\documentclass[useAMS,usegraphicx]{mn2e}

\usepackage{times}

\newif\ifAMStwofonts

\newcommand{\kms}{\ensuremath{\hbox{km}\cdot \hbox{s}^{-1}}}
\newcommand{\hmpc}{\ensuremath{h^{-1}\,\hbox{Mpc}}}
\newcommand{\mhmpc}{{\, h^{-1}\rm Mpc}}

\newcommand{\de}{\delta}
\newcommand{\te}{\theta}

\newcommand{\f}{\frac}

\newcommand{\s}{\sigma}

\newcommand{\bfx}{{\bf x}}
\newcommand{\bfr}{{\bf r}}

\newcommand{\bfy}{{\bf y}}
\newcommand{\bfk}{{\bf k}}
\newcommand{\bfv}{{\bf v}}

\newcommand{\bfg}{{\bf g}}

\newcommand{\calO}{{\cal O}}

\newcommand{\calR}{{\cal R}}
\newcommand{\eps}{{\epsilon}}
\newcommand{\bc}{\begin{center}}
\newcommand{\be}{\begin{equation}}
\newcommand{\ee}{\end{equation}}
\newcommand{\ec}{\end{center}}
\newcommand{\lan}{\langle}
\newcommand{\ran}{\rangle}
\newcommand{\mcL}{\mathcal{L}}
\newcommand{\mcR}{\mathcal{R}}

\newcommand{\spose}[1]{\hbox to 0pt{#1\hss}}
\newcommand{\lta}{\mathrel{\spose{\lower 3pt\hbox{$\mathchar"218$}}
 \raise 2.0pt\hbox{$\mathchar"13C$}}}
\newcommand{\gta}{\mathrel{\spose{\lower 3pt\hbox{$\mathchar"218$}}
 \raise 2.0pt\hbox{$\mathchar"13E$}}}

\newcommand{\etal}{{et al.}~}
\newcommand{\err}{r}

\begin{document}

\title[]{Likelihood analysis of the Local Group acceleration revisited}

\author[Cieciel\c{a}g \& Chodorowski]
{Pawe{\l} Cieciel\c{a}g$^{1,2}$\thanks{E-mail: pci@usm.uni-muenchen.de}
and Micha{\l} J.\ Chodorowski$^2$\thanks{E-mail: michal@camk.edu.pl}\\
$^1$Institute of Astronomy and Astrophysics, Munich University, 
Scheinerstr. 1, D-81679 Munich, Germany \\
$^2$Copernicus Astronomical Center, Bartycka 18, 00--716 Warsaw,
Poland }

\maketitle
\begin{abstract}
We reexamine likelihood analyzes of the Local Group (LG) acceleration,
paying particular attention to nonlinear effects. Under the
approximation that the joint distribution of the LG acceleration and
velocity is Gaussian, two quantities describing nonlinear effects
enter these analyzes. The first one is the coherence function,
i.e. the cross-correlation coefficient of the Fourier modes of gravity
and velocity fields. The second one is the ratio of velocity power
spectrum to gravity power spectrum. To date, in all analyzes of the LG
acceleration the second quantity was not accounted for. Extending our
previous work, we study both the coherence function and the ratio of
the power spectra. With the aid of numerical simulations we obtain
expressions for the two as functions of wavevector and
$\sigma_8$. Adopting WMAP's best determination of $\sigma_8$, we
estimate the most likely value of the parameter $\beta$ and its
errors. As the observed values of the LG velocity and gravity, we
adopt respectively a CMB-based estimate of the LG velocity, and
Schmoldt et al.'s (1999) estimate of the LG acceleration from the PSCz
catalog. We obtain $\beta = 0.66^{+0.21}_{-0.07}$; thus our errorbars
are significantly smaller than those of Schmoldt et al. This is not
surprising, because the coherence function they used greatly
overestimates actual decoherence between nonlinear gravity and
velocity.
\end{abstract}



\section{Introduction}
\label{sec:intro}

Analyzes of large-scale structure of the Universe provide estimates of
cosmological parameters that are complementary to those from the
cosmic microwave background (CMB) measurements. In particular,
comparing the large-scale distribution of galaxies to their peculiar
velocities enables one to constrain the quantity $\beta \equiv
\Omega_m^{0.6}/b$. Here, $\Omega_m$ is the cosmological matter density
parameter and $b$ is the linear bias of galaxies that are used to
trace the underlying mass distribution. This is so because the
peculiar velocity field, $\bfv$, is induced gravitationally and
therefore is tightly coupled to the matter distribution. In the linear
regime, this relationship takes the form

\be
\bfv = \Omega_m^{0.6} \int \frac{{\rm d}^3\bfr}{4 \pi} 
\delta(\bfr) \frac{\bfr}{r^3} \,,
\label{eq:lin_vel}
\ee
where $\de$ denotes the mass density contrast and distances have been
expressed in \kms. Under the assumption of linear bias $\de = b^{-1}
\de_g$, where $\de_g$ denotes the density contrast of galaxies, and
the amplitude of peculiar velocities depends linearly on $\beta$.

These comparisons are done by extracting the density field from
full-sky redshift surveys (such as the PSCz; Saunders \etal 2000), and
comparing it to the observed velocity field from peculiar velocity
surveys.  The methods for doing this fall into two broad categories.
One can use equation~(\ref{eq:lin_vel}) to calculate the predicted
velocity field from a redshift survey, and compare the result with the
measured peculiar velocity field; this is referred to as a
velocity--velocity comparison. Alternatively, one can use the
differential form of this equation, and calculate the divergence of
the observed velocity field to compare directly with the density field
from a redshift survey; this is called a density--density comparison.

Peculiar velocities of galaxies and groups of galaxies are generally
determined by measuring their distances independently of
redshifts. However, the motion of the Local Group (LG) of galaxies can
be deduced in another way, namely from the observed dipole anisotropy
of the CMB temperature. This dipole reflects, via the Doppler shift,
the motion of the Earth with respect to the CMB rest frame. The
components of this motion of local, non-cosmological origin (the Earth
motion around the Sun, the Sun motion in the Milky Way and the motion
of the Milky Way in the LG) are known and can be subtracted (e.g.,
Courteau \& van den Bergh 1999). When transformed to the barycenter of
the LG, the motion is towards $(l,b) = (276^{\circ} \pm
3^{\circ},30^{\circ} \pm 2^{\circ})$, and of amplitude $v_{\rm LG} =
627 \pm 22$ \kms, as inferred from the 4-year COBE data (Lineweaver
\etal 1996). It can be compared to that predicted from a redshift 
survey and historically this was the first velocity--velocity
comparison (with angular data only, e.g. Meiksin \& Davis 1986, Yahil,
Walker \& Rowan-Robinson 1986; with redshift information, Strauss \&
Davis 1988, Lynden-Bell, Lahav \& Burstein 1989). This velocity
estimate is much more accurate than estimates of peculiar velocities
based on redshift-independent distances. Therefore, the analysis of
the LG motion remains an interesting alternative to current
velocity--velocity comparisons, performed simultaneously for many
galaxies, but with less accurately measured velocities.

Let us define the {\em scaled\/} gravity, $\bfg$, by the equation:

\be
\bfg = \int \frac{{\rm d}^3\bfr}{4 \pi} 
\delta(\bfr) \frac{\bfr}{r^3} \,,
\label{eq:grav}
\ee
where again distances have been expressed in \kms. The scaled gravity
is {\em proportional\/} to the gravitational acceleration, and can be
measured from a redshift survey. Equation~(\ref{eq:lin_vel}) yields

\be
\bfv = \Omega_m^{0.6} \bfg \,,
\label{eq:v--g}
\ee
so a velocity--velocity comparison is in fact a velocity--gravity 
one. Hereafter, we will refer to `scaled gravity' simply by
`gravity'. It is also often called `the clustering dipole'.

A number of effects (nonlinear effects, observational windows through
which the velocity and gravity of the LG are observed, shot noise)
spoil the linear relationship~(\ref{eq:v--g}). Therefore, to estimate
$\beta$ one cannot simply equate the two dipoles, but a more
sophisticated approach is needed. Commonly adopted is a maximum
likelihood approach, which provides a way to account for these effects
and to compute errors of estimated cosmological parameters. Here we
reexamine a likelihood analysis of Schmoldt \etal (1999; S99), paying
particular attention to proper modelling of nonlinear effects (NE),
that affect the estimated values of parameters and their errors.

In a previous work (Chodorowski \& Cieciel\c ag 2002; hereafter C02)
we concentrated on one quantity describing NE in the LG
velocity--gravity comparison, the coherence function (CF). It is the
cross-correlation coefficient of the Fourier modes of the gravity and
velocity fields. We showed that the form of the function adopted by
S99 drastically overestimates actual decoherence between nonlinear
gravity and velocity. This implies that the true random error of
$\beta$ is significantly smaller. In the present study we give a
complete description of NE present in the analysis. In particular, we
show that not only the CF is relevant, but also the ratio of the power
spectrum of velocity to the power spectrum of gravity. With aid of
numerical simulations we model the two quantities as functions of the
wavevector and of cosmological parameters. We then combine these
results with observational estimates of $\bfv_{\rm LG}$ and $\bfg_{\rm
LG}$, and obtain the `best' value of $\beta$ and its errors. The paper
is organized as follows. In Section~\ref{sec:anal} we outline an
analytical description of the likelihood for cosmological parameters,
based on the measured values of the LG velocity and acceleration. In
Section~\ref{sec:num} we describe our numerical simulations, and we
model numerically the coherence function and the ratio of the power
spectra. An estimation of the most likely value of $\beta$ and its
errors is presented in Section~\ref{sec:param}. Summary and
conclusions are in Section~\ref{sec:conc}.

\section{Analytical description of the likelihood}
\label{sec:anal} 

Let $f(\bfg,\bfv)$ denote the joint distribution function for the LG
gravity and velocity. It is commonly approximated by a multivariate
Gaussian (Strauss \etal 1992, hereafter S92; S99). This approximation
has support from numerical simulations (Kofman \etal 1994, Cieciel{\c
a}g \etal 2003), where the measured nongaussianity of $\bfg$ and
$\bfv$ is small. This is rather natural to expect since, e.g., gravity
is an integral of density over effectively a large volume, so the
central limit theorem can at least partly be applicable (but see
Catelan \& Moscardini 1994). 

In a Bayesian approach, one ascribes {\it a priori} equal
probabilities to values of unknown parameters, which allows one to
express their likelihood function via $f$:
\begin{equation}
\label{eq:Bayes}
\mcL(\rm param.) = f(\bfv,\bfg~|~\rm param.) \,.
\end{equation}
As the parameters to be estimated we adopt $\beta$ and $b$. Using
statistical isotropy of $\bfg$ and $\bfv$, the distribution function
can be cast to the form (Juszkiewicz \etal 1990, Lahav, Kaiser \&
Hoffman 1990):

\be
f(\bfg,\bfv) = \f{(1 - \err^2)^{-3/2}}{(2 \pi)^{3} \s_\bfg^{3} 
\s_\bfv^{3}} 
\exp\left[- \f{x^2 + y^2 - 2 \err \mu x y}{2(1 - \err^2)}\right] \,,
\label{eq:dist}
\ee
where $\s_\bfg$ and $\s_\bfv$ are the r.m.s.\ values of a single
Cartesian component of gravity and velocity, respectively. From
isotropy, $\s_\bfg^2 = \lan \bfg\cdot\bfg \ran/3$, and $\s_\bfv^2 =
\lan \bfv\cdot\bfv\ran/3$, where $\lan \cdot \ran$ denote the ensemble
averaging. Next, $(\bfx,\bfy) = (\bfg/\s_\bfg,\bfv/\s_\bfv)$, and $\mu
= \cos\psi$ with $\psi$ being the misalignment angle between
$\bfg$ and $\bfv$. Finally, $\err$ is the cross-correlation
coefficient of $g_m$ with $v_m$, where $g_m$ ($v_m$) denotes an
arbitrary Cartesian component of $\bfg$ ($\bfv$). From isotropy,

\be
\err = \f{\lan \bfg \cdot \bfv \ran}{\lan g^2 \ran^{1/2} \lan v^2
\ran^{1/2}} \,.
\label{eq:err}
\ee
Also from isotropy,
\be
\lan x_m y_n \ran = \err \, \de_{mn} \,,
\label{eq:cross}
\ee
where $\de_{mn}$ denotes the Kronecker delta. In other words, there
are no cross-correlations between different spatial components. 

We have to take into account that the LG gravity is measured through
the window $W_\bfg$:

\be
\bfg = \int \f{{\rm d}^3 r}{4 \pi} \de(\bfr) W_\bfg(r) \f{\bfr}{r^3}
\,.
\label{eq:gint}
\ee
If $\bfv$ is irrotational like $\bfg$, then similarly

\be
\bfv = \int \f{{\rm d}^3 r}{4 \pi} \te(\bfr) W_\bfv(r) \f{\bfr}{r^3}
\,,
\label{eq:vint}
\ee
where $\te$ is the (minus) velocity divergence, $\te \equiv - \nabla
\cdot \bfv$, and $W_\bfv$ is the velocity window. Due to Kelvin's
circulation theorem, the cosmic velocity field is vorticity-free as
long as there is no shell crossing.  N-body simulations (Bertshinger
\& Dekel 1989, Mancinelli et al. 1994, Pichon \& Bernardeau 1999) 
show that the vorticity of velocity is small in comparison to its
divergence even in the fully nonlinear regime.

The best current estimate of the LG gravity is inferred from the PSCz
catalog of {\em IRAS} galaxies (Saunders \etal 2000). S99 follow S92
and measure the LG gravity through the standard {\em IRAS} window,

\be
W_\bfg = \left\{ \begin{array}{ll} (r/r_s)^3, & r < r_s \,, \\ 1, &
r_s < r < R_{\rm max} \,, \\ 0, & R_{\rm max} < r \,.
\end{array} \right. 
\label{eq:W_g}
\ee
The window is characterized by a small-scale smoothing and a sharp
large-scale cutoff. Following S99, we adopt the values $r_s = 500$
\kms\ and $R_{\rm max} = 15,000$ \kms, appropriate for the PSCz 
catalog. The window function relevant to the LG velocity is

\be
W_\bfv = \left\{ \begin{array}{ll}
0, & r < r_{\rm min} \,, \\
1, & \mbox{otherwise}\,,
\end{array} \right. 
\label{eq:W_v}
\ee
which has a small-scale cutoff, $r_{\rm min} = 100$ \kms, to reflect
the finite size of the LG (S92). Using numerical simulations, we have
checked that the velocity field remains approximately Gaussian even
for such a small smoothing scale.

In Fourier space, relations~(\ref{eq:gint}) and~(\ref{eq:vint}) read:
\begin{equation}
\label{eq:g_Four}
\bfg_{\bfk}=\frac{i\bfk}{k^2}\delta_{\bfk}\widehat{W}_g(k),
\end{equation}
\begin{equation}
\label{eq:v_Four}
\bfv_{\bfk}=\frac{i\bfk}{k^2}\theta_{\bfk}\widehat{W}_v(k),
\end{equation}
where the subscript $\bfk$ denotes the Fourier transform. The quantity
$\widehat{W}$ is not a Fourier transform of $W$, but is related to it
in the following way (S92):

\begin{equation}
\label{eq:k_okno}
\widehat{W}(k) \equiv k \!\int_0^\infty \! W(r) j_1(kr) dr \,.
\end{equation}
Here and below $j_l$ represents the spherical Bessel function of
order $l$. In particular, 

\be
\widehat{W}_g(k) = \f{3 j_1(k r_s)}{k r_s} - j_0(k R_{\rm max}) \,,
\label{eq:W_g(k)}
\ee
and

\be
\widehat{W}_v(k) = j_0(k r_{\rm min}) \,.
\label{eq:W_v(k)}
\ee

From equations~(\ref{eq:g_Four}) and~(\ref{eq:v_Four}) we have

\be
\lan \bfg \cdot \bfg \ran = \f{1}{2 \pi^2} \int_0^\infty
\widehat{W}_\bfg^2(k) P(k) dk \,,
\label{eq:g^2}
\ee
and

\be
\lan \bfv \cdot \bfv \ran = \f{1}{2 \pi^2} \int_0^\infty
\widehat{W}_\bfv^2(k) P_{\te}(k) dk \,.
\label{eq:v^2_1}
\ee
Here, $P(k)$ and $P_{\te}(k)$ are respectively the power spectrum of
the density and the power spectrum of the velocity divergence. 
Thus,

\be
\lan \bfv \cdot \bfv \ran = \f{1}{2 \pi^2} \int_0^\infty
\widehat{W}_\bfv^2(k) \calR(k) P(k) dk \,,
\label{eq:v^2}
\ee
where

\be
\calR(k) \equiv \f{P_{\bfv}(k)}{P_{\bfg}(k)} = 
\f{P_{\te}(k)}{P(k)} \,,
\label{eq:calR}  
\ee
and $P_{\bfv}$ and $P_{\bfg}$ are the power spectra respectively of
velocity and gravity. Furthermore,

\be
\lan \bfg \cdot \bfv \ran = \f{1}{2 \pi^2} \int_0^\infty
\widehat{W}_\bfg(k) \widehat{W}_\bfv(k) C(k) 
P_{\te}^{1/2}(k) P^{1/2}(k) dk ,
\label{eq:g-v}
\ee
where $C(k)$ is the so-called {\em coherence function}\footnote{S92
call it the {\em decoherence} function. We prefer the name
`coherence', because higher values of the function imply higher, not
lower, correlation between gravity and velocity.} (S92), or the
correlation coefficient of the Fourier components of the gravity and
velocity fields:

\be
C(k) \equiv \f{\lan \bfg_\bfk \cdot \bfv_\bfk^\star \ran}{\lan
|\bfg_\bfk|^2 \ran^{1/2} \lan |\bfv_\bfk|^2 \ran^{1/2}} 
= \f{\lan \de_\bfk \te_\bfk^\star \ran}{\lan |\de_\bfk|^2 
\ran^{1/2} \lan |\te_\bfk|^2 \ran^{1/2}}
\,.
\label{eq:coh_def} 
\ee
Hence, we obtain
\be 
\err = \f{\int_0^\infty \widehat{W}_\bfg(k) \widehat{W}_\bfv(k)
C(k) \calR^{1/2}(k) P(k) dk}{\left[\int_0^\infty \widehat{W}_\bfg^2(k) 
P(k) dk\right]^{1/2} \left[\int_0^\infty \widehat{W}_\bfv^2(k) 
\calR(k) P(k) dk\right]^{1/2}} \,.
\label{eq:err_form} 
\ee

Equations~(\ref{eq:g^2}), (\ref{eq:v^2}) and~(\ref{eq:err_form})
specify all parameters (the variances and the correlation coefficient)
that determine distribution~(\ref{eq:dist}) {\em in the absence of
observational errors}. The deviation of the correlation coefficient
from unity is then due to different windows, through which the gravity
and the velocity of the LG are measured, and due to nonlinear
effects. The latter are described by two functions: the coherence
function, and the ratio of the power spectra. In the linear regime,
$C(k) = 1$ and $\calR = \Omega_m^{1.2}$. This yields

\be 
\err = \f{\int_0^\infty \widehat{W}_\bfg(k) \widehat{W}_\bfv(k)
P(k) dk}{\left[\int_0^\infty \widehat{W}_\bfg^2(k) 
P(k) dk\right]^{1/2} \left[\int_0^\infty \widehat{W}_\bfv^2(k) P(k)
dk\right]^{1/2}} \,,
\label{eq:err_form_2} 
\ee
so for linear fields the correlation coefficient is determined solely
by the windows, as expected.

\section{Modelling nonlinear effects}
\label{sec:num}

\subsection{Numerical simulations}
We follow the evolution of the dark matter distribution using the
pressureless hydrodynamic code CPPA (Cosmological Pressureless
Parabolic Advection, see Kudlicki, Plewa \& R\'o\.zyczka 1996,
Kudlicki \etal 2000 for details). It employs an Eulerian scheme with
third order accuracy in space and second order in time, which assures
low numerical diffusion and an accurate treatment of high density
contrasts. Standard applications of hydrodynamic codes involve a
collisional fluid; however, we implemented a simple flux interchange
procedure to mimic collisionless fluid behaviour. Thanks to this
approach we avoid a few problems of N-body codes. The main advantage
of a hydrodynamic code over an N-body code is accurate treatment of
low density regions. Moreover it directly produces a volume-weighted
velocity field.  This is important because in the definition of the
CF, equation~(\ref{eq:coh_def}), the velocity field is
volume-weighted, not mass-weighted. Furthermore, the Fast Fourier
Transform can be directly applied to the data on an uniform grid.

The computational domain forms a $(400\mhmpc)^3$ cube with $256^3$
grid cells and periodic boundary conditions.  This setup allows us to
cover a broad range of wavenumbers, $k\in[0.016,2.]$ $h/\hbox{Mpc}$,
which is important for accurate calculation of the integrals in
equation (\ref{eq:err_form}). The initial distribution of the density
fluctuations is Gaussian and their power spectrum is given by a CDM
formula (as in Eq.~7 of Efstathiou, Bond \& White 1992) with the shape
parameter $\Gamma=0.19$, as inferred from the {\em IRAS\/} PSCz survey
(Sutherland \etal 1999) and in agreement with recent determinations
from WMAP (Spergel \etal 2003). The initial value of the square root
of the variance of the density field in spheres of radius 8
\hmpc, $\s_{8}(t_i)$, is $0.019$. We studied a range of outputs from
the simulations, corresponding to different values of
$\s_8$. Specifically, full output data were dumped in constant
intervals of the scale factor, $\Delta a = 0.025$.

All our simulations assume $\Omega_m = 0.3$, $\Omega_\Lambda = 0$. In
the mildly non-linear regime, the quantity $\tilde\bfv \equiv
\Omega_m^{-0.6} \bfv$ is insensitive to the cosmological density
parameter and cosmological constant, as demonstrated both analytically
(Bouchet \etal 1995, Nusser \& Colberg 1998; see also Appendix B3 of
Scoccimarro \etal 1998) and by means of N-body simulations (Bernardeau
\etal 1999). Therefore, our results should be valid for any
cosmology. Specifically, we have

\be
\calR = \Omega_m^{1.2} \tilde\calR 
\label{eq:scaled_R} 
\ee
and

\be
C = \tilde C \,,
\label{eq:scaled_C} 
\ee
where  

\be
\tilde\calR(k) = \f{P_{\tilde{\bfv}}(k)}{P_{\bfg}(k)} 
\label{eq:scaled_calR}  
\ee
and

\be
\tilde C(k) = \f{\lan \bfg_\bfk \cdot \tilde \bfv_\bfk^\star 
\ran}{\lan|\bfg_\bfk|^2 \ran^{1/2} 
\lan |\tilde\bfv_\bfk|^2 \ran^{1/2}} \,.
\label{eq:scaled_coh} 
\ee
In a previous paper (C02), we tested numerically the dependence of the
coherence function on $\Omega_m$ and found it to be extremely weak.

In C02, we also investigated numerical effects. In short, we found
that the effects of resolution affect numerical determination of the
CF at scales smaller than 4 grid cells, ie. twice the Nyquist
wavelength. Therefore, all the results we present here are for $k<1$
$h/\hbox{Mpc}$. For these wavenumbers, both simulated $C$ and $\calR$
practically do not depend on resolution. Also, here we use a larger
box size than in C02 in order to model longer modes, while the short
wavelength limit ($k<1$ $h/\hbox{Mpc}$) still allows us to accurately
calculate the integrals in Equation~(\ref{eq:err_form}), as will be
shown in subsection~\ref{sec:converg}.

\subsection{Coherence function}
\label{sec:coher}
\begin{figure}
\centerline{\includegraphics[angle=0,scale=0.36]{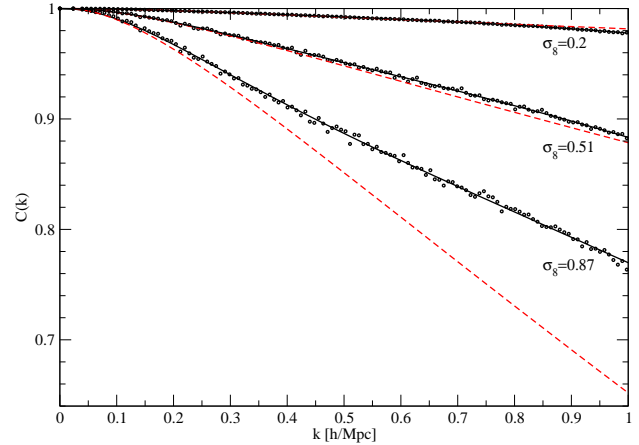}}
\caption{\label{fig:CF} The coherence function for various values of 
$\sigma_8$. Dots show the results of simulations, solid lines -- our
fit (\ref{eq:CF_fit}), and dashed lines -- the results of perturbative
calculations. Perturbative approximation breaks down for $\s_8 > 0.5$.}
\end{figure}

We studied the CF in C02. We found there that the characteristic
decoherence scale is an order of magnitude smaller than previously
used (S92). The weak point of the formula fitted in C02 was a poor
description of the CF for long-wavelength modes. These modes are
however important when calculating integrals in
equation~(\ref{eq:err_form}). (From
eqs.~\ref{eq:g_Four}--\ref{eq:v_Four} it follows that the gravity and
velocity fields are more sensitive to long-wavelength modes than the
density field.) Therefore, here we use another fitting function, which
is more accurate for low values of $k$:

\begin{equation}
\label{eq:CF_fit}
\tilde C(k) = \left [ 1+(a_0 k - a_2 k^{1.5} + a_1 k^2)^{2.5} 
\right ]^{-0.2}.
\end{equation}
Parameters $a_l$ were obtained for 35 different values of $\sigma_8$
in the range [0.1,1], and we found the following, power-law, scaling
relations:
\begin{eqnarray}
a_0 &=& 4.908\ \sigma_8^{0.750} \nonumber \\
a_1 &=& 2.663\ \sigma_8^{0.734} \\
a_2 &=& 5.889\ \sigma_8^{0.714} \nonumber .
\end{eqnarray}
The fit was calculated for $k\in [0, 1]$ $h/\hbox{Mpc}$, with the
imposed constraint $C(k=0)=1$. Unlike the previous fit, the new one
results in the value of the correlation coefficient of gravity and
velocity which agrees with that measured directly in our
simulations. In Figure \ref{fig:CF} we show, for various values of
$\sigma_8$, the CF from simulations, its fit (\ref{eq:CF_fit}), and
results of perturbative calculations, described in C02. We see that
the perturbative approximation breaks down for $\sigma_8 > 0.5$.

\subsection{Ratio of the power spectra}
\label{sec:vgpower}

The ratio of the velocity to the gravity power spectra, $\mcR$, is
related to its scaled counterpart, $\tilde\calR$, by
equation~(\ref{eq:scaled_R}). The quantity $\tilde\calR$, defined in
equation~(\ref{eq:scaled_calR}), practically does not depend on the
background cosmological model. It departs from unity in the nonlinear
regime because the velocity grows slower than it would be expected
from the linear approximation. 

\begin{figure}
\centerline{\includegraphics[angle=-90,scale=0.3]{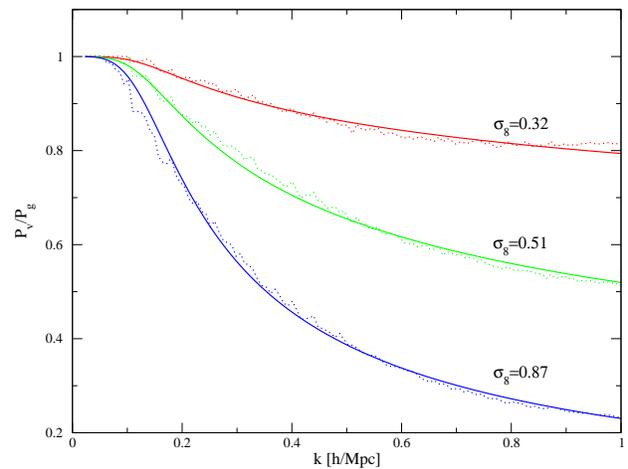}}
\caption{\label{fig:R} The ratio of the velocity power spectrum to the
gravity power spectrum as a function of the wavevector for three
different values of $\sigma_8$. Dotted lines show the results of our
simulations and solid lines -- the fit~(\ref{eq:pvpgfit_fix}).}
\end{figure}

We have found that $\tilde\mcR$ obtained from simulation can be fitted
with the following formula:

\be
\label{eq:pvpgfit_fix}
\tilde\mcR(k) = [1+(7.071k)^4]^{-\alpha} \,,
\ee
with

\be 
\alpha = -0.06574 + 0.29195\sigma_8 \qquad 
\mathrm{for}~0.3<\sigma_8<1 \,.
\ee
We stress that the above relation between $\alpha$ and $\sigma_8$ is
valid for $\sigma_8\in[0.3,1]$, which is still a sufficiently wide
range of values. Figure \ref{fig:R} shows the ratio of the power
spectra from simulation, and our fit.

\subsection{Convergence of $\bmath{r}$}
\label{sec:converg}

We calculate the correlation coefficient, $r$, by inserting fits
(\ref{eq:CF_fit}) and (\ref{eq:pvpgfit_fix}) into
formula~(\ref{eq:err_form}). However, the integrals in this formula
extend over the whole $k$-space, while the fits have been obtained for
a limited range of wavenumbers between $0.016$ and $1$ $h/\hbox{Mpc}$.
Therefore, the question has to be answered whether this extrapolation
is justified.

\begin{figure}
\centerline{\includegraphics[angle=0,scale=0.45]{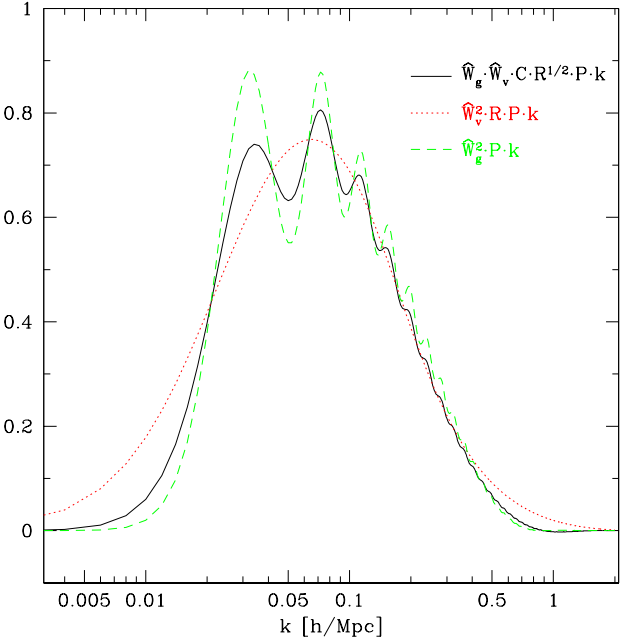}}
\caption{\label{fig:pwin} Integrands of the integrals appearing in 
equation~(\ref{eq:err_form}) as functions of the wavevector for $\sigma_8=0.84$. 
Since the $k$-axis is logarithmic, the ordinate is multiplied by and extra power
of $k$, so equal areas under a function correspond to equal
contributions to an integral. Solid line shows the product
$\widehat{W}_\bfg \widehat{W}_\bfv C \calR^{1/2} P k$, dotted --
$\widehat{W}_\bfv^2 \calR P k$, and dashed -- $\widehat{W}_\bfg^2 P
k$.  Units of the ordinate axis are arbitrary.}
\end{figure}

Figure~\ref{fig:pwin} shows integrands of the integrals in the
formula~(\ref{eq:err_form}) for $\sigma_8=0.84$. It is evident that
contributions from wavenumbers greater than unity are negligible. This
is so because observational windows of the LG gravity and velocity
filter out smaller scales.\footnote{Strictly speaking, the velocity
window passes contributions from wavenumbers up to about $2$
$h/\hbox{Mpc}$, but the ratio of the power spectra damps them
additionally for $k>1$ $h/\hbox{Mpc}$.} This is not quite the case for
wavenumbers smaller than $0.016$ $h/\hbox{Mpc}$, but these scales are
well within the linear regime, for which the limiting values of the CF
and the ratio of the power spectra are known to converge to unity.

\section{Parameter estimation}
\label{sec:param}
We are now applying our formalism to the PSCz survey. As the value of
$\s_8$ we adopt its WMAP's estimate, $\s_8 = 0.84$ ($\pm 0.04$;
Spergel et al. 2003). This specifies the coherence function and the
ratio of the power spectrum of velocity to the power spectrum of
gravity. Also, this provides a normalization for the power spectrum of
density.

\subsection{Parameter dependence of the model}
As stated before, the likelihood of specific values of $\beta$ and $b$
is determined by the distribution~(\ref{eq:dist}). In this
distribution, the observables are $\bfg$ and $\bfv$, or $g$, $v$, and
the misalignment angle, $\psi$. Following S99, we adopt for them the
following values: $g = 933 \;\kms$ (from the distribution of the PSCz
galaxies up to $150\;\hmpc$), $v = 627\; \kms$ (inferred from the
4-year COBE data by Lineweaver \etal 1996), and $\psi = 15^\circ$.

The theoretical quantities are $\s_\bfg$, $\s_\bfv$, and $\err$. The
variance of a single spatial component of measured gravity,
$\s_\bfg^2$, is a sum of the cosmological component, $\s_{\bfg,c}^2$,
and errors, $\eps^2$. Since gravity here is inferred from a galaxian,
rather than mass, density field, we have $\s_{\bfg,c}^2 = b^2
s_\bfg^2$, where $s_\bfg^2$ is the variance of a single component of
the true (i.e., mass-induced) gravity. From equation~(\ref{eq:g^2}),

\begin{equation}
s_\bfg^2 = \frac{1}{6\pi^2}\int_0^{\infty} \widehat{W}_g^2(k) P(k) dk
\,.
\label{eq:s_g}
\end{equation}

The gravity errors are twofold: due to finite sampling of the galaxy
density field, and due to the reconstruction of the galaxy density
field in real space. Therefore, $\eps^2 = (\s_{\rm SN}^2 + \s_{\rm
rec}^2)/3$, where $\s_{\rm SN}^2$ and $\s_{\rm rec}^2$ are
respectively the shot noise (or, sampling) variance and the
reconstruction variance. (Both $\s_{\rm SN}^2$ and $\s_{\rm rec}^2$
are full, i.e., 3D, variances.) Estimated by S99 using mock catalogs,
the cumulative shot noise at 150 \hmpc\ amounts to $\s_{\rm SN} = 160$
\kms. An average reconstruction error in the differential contribution
to the cumulative gravity, produced by a shell of matter $10$ \hmpc\
wide, is $15$ \kms. Since up to $150$ \hmpc\ there are $15$ such
shells, we have $\s_{\rm rec} =
\sqrt{15}\cdot 15$ \kms $= 58$ \kms. To sum up,

\be
\s_\bfg^2 = b^2 s_\bfg^2 + \f{\s_{\rm SN}^2 + \s_{\rm rec}^2}{3} \,,
\label{eq:sigma_g}
\ee
where

\be
\s_{\rm SN} = 160\; \kms \qquad {\rm and} \qquad 
\s_{\rm rec} = 58\; \kms \,.
\label{eq:sigma_g_errors}
\ee
 
Errors in the measured velocity of the LG are negligible compared to
those in the gravity. Equations~(\ref{eq:v^2}) and~(\ref{eq:scaled_R})
yield 

\be
\s_\bfv =  \Omega_m^{0.6} s_\bfv \,,
\label{eq:sigma_v}
\ee 
where 

\be
s_\bfv^2 = \frac{1}{6\pi^2}\int_0^{\infty} \widehat{W}_v^2(k)
\tilde\mcR(k) P(k) dk \,.
\label{eq:s_v}
\end{equation}
Finally, errors in the estimate of the LG gravity do not affect the
cross-correlation between the LG gravity and velocity, but increase
the gravity variance. This has the effect of lowering the value of the
cross-correlation coefficient. Specifically, from
equations~(\ref{eq:scaled_R})--(\ref{eq:scaled_C}) and
(\ref{eq:sigma_g}) we have

\be
\err = \rho \left(1+ \f{\s_{\rm SN}^2 + 
\s_{\rm rec}^2}{3 b^2 s_\bfg^2} \right)^{-1/2} ,
\label{eq:err_errors} 
\ee 
where

\be
\rho = \f{\int_0^\infty \widehat{W}_\bfg(k) \widehat{W}_\bfv(k)
\tilde{C}(k) \tilde\calR^{1/2}(k) P(k) dk}{\left[\int_0^\infty 
\widehat{W}_\bfg^2(k) P(k) dk\right]^{1/2} \left[\int_0^\infty 
\widehat{W}_\bfv^2(k) \tilde\calR(k) P(k) dk\right]^{1/2}} \,.
\label{eq:scaled_err} 
\ee 
Thus, the likelihood depends explicitly on the parameters $b$ and
$\Omega_m$. However, for reasons that will become evident later on, as
the parameters to be estimated we choose $b$ and $\beta \equiv
\Omega_m^{0.6}/b$. Since $\Omega_m^{0.6} = \beta b$, both $\s_\bfg$, 
$\s_\bfv$ and $r$ depend on $b$. On the other hand, only $\s_\bfv$
depends on $\beta$. This makes the derivation of the most likely value
of $\beta$, given $b$, simple.

\subsection{$\bmath{\beta}$ for known $\bmath{b}$}
Using equations~(\ref{eq:dist}),~(\ref{eq:sigma_v}), and the equality
$\Omega_m^{0.6} = \beta b$, the logarithmic likelihood for $\beta$
takes on the form:
\begin{eqnarray}
\ln \mcL(\beta) 
\!\!\!\! &=& \!\!\!\! - 3 \ln{2\pi} - 3 \ln[\s_\bfg b s_\bfv 
(1-r^2)^{1/2}] - 3 \ln \beta \nonumber \\ 
\!\!\!\! & & \!\!\!\! - \frac{1}{2(1-r^2)}\left (
\frac{g^2}{\s_\bfg^2} + \frac{v^2}{\beta^2 b^2 s_\bfv^2} - 
\frac{2r\mu g v}{\s_\bfg \beta b s_\bfv} \right ).
\end{eqnarray}
To find its maximum, we calculate its partial derivative with respect
to $\beta$, $\partial \ln \mcL/\partial \beta$, and equate it to
zero. This yields the following equation:

\begin{equation}
\label{eq:kwadrat}
3(1-r^2)\beta^2 + \frac{r\mu g v}{\s_\bfg b s_\bfv} \beta -
\frac{v^2}{b^2 s_\bfv^2} = 0 \,.
\end{equation}
The LG gravity, inferred from the PSCz survey, is tightly coupled to
its velocity, $1 - \err \ll 1$ and $1 - \mu \ll 1$. (Specifically,
$\mu = 0.97$ and, for $\s_8$ around $0.8$, $\err \simeq 0.93$.) At
first approximation we can therefore assume $\err = \mu = 1$, hence

\begin{equation}
\label{eq:beta1}
\beta_1 = \frac{\s_\bfg}{b s_\bfv} \frac{v}{g} \,.
\end{equation}
Using equation~(\ref{eq:sigma_g}) we obtain finally

\be
\label{eq:beta1_fin}
\beta_1 = \frac{s_\bfg}{s_\bfv} \left(1+ \f{\s_{\rm SN}^2 + 
\s_{\rm rec}^2}{3 b^2 s_\bfg^2} \right)^{1/2} \frac{v}{g} \,.
\ee
Thus, the best estimate of $\beta$ is not just the ratio of the LG
velocity to its gravity: it is modified by nonlinear effects (which
affect $s_\bfv$ through the function $\tilde\calR$), different
observational windows (which affect differently $s_\bfg$ and
$s_\bfv$), and observational errors.

At next approximation, in equation~(\ref{eq:kwadrat}) one could
approximate $\beta^2$ by $\beta_1^2$. However, we have considered the
case of known $b$ for illustrative purposes only and from now on we
relax this assumption. In the next subsection we will analyze the
joint likelihood for $\beta$ and $b$.

\subsection{Joint likelihood for $\bmath{\beta}$ and $\bmath{b}$} 

\begin{figure}
\centerline{\includegraphics[angle=0,scale=0.5]{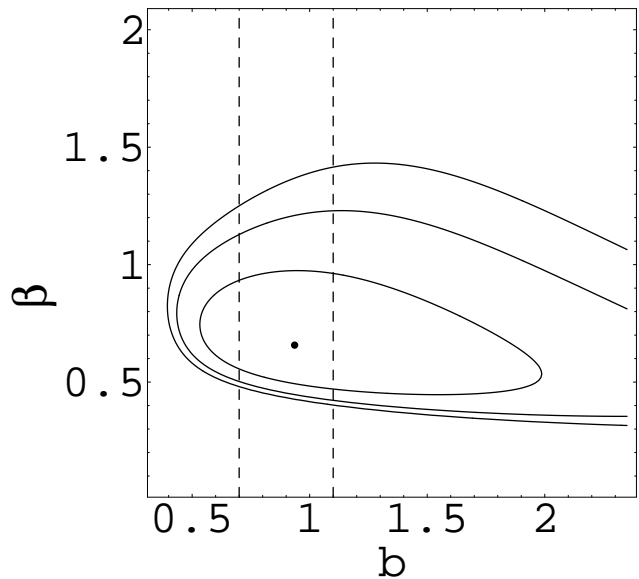}}
\caption{\label{fig:lh} Likelihood contours for the parameters $\beta$
and $b$, corresponding to the confidence levels of 68, 90 and
95\%. The maximum of the likelihood function is denoted by a dot.}
\end{figure}

Figure~\ref{fig:lh} shows isocontours of the joint likelihood for
$\beta$ and $b$, corresponding to the confidence levels of 68, 90 and
95\%. The maximum of the likelihood is denoted by a dot. The
corresponding values of $\beta$ and $b$ are respectively $0.66$ and
$0.94$. 

This figure illustrates the well-known problem of $\Omega_m$--bias
degeneracy in cosmic density--velocity comparisons. A `common wisdom'
is that in these comparisons, only a degenerate combination of
$\Omega_m$ and $b$, $\beta \equiv \Omega_m^{0.6}/b$, can be
determined. Here this is not strictly true, since we have estimated
the most likely values of both $\beta$ and $b$, so in principle we
could solve for the best value of $\Omega_m$ alone. However, the
isocontours of the likelihood are much more elongated along the
$b$-axis than along the $\beta$-axis, making the resulting constraints
on $\Omega_m$ much weaker than on $\beta$. In practice, therefore,
from our analysis we cannot say much about $\Omega_m$ or bias
separately. On the other hand, we can put fairly tight constraints on
$\beta$.

\begin{figure}
\centerline{\includegraphics[angle=0,scale=0.5]{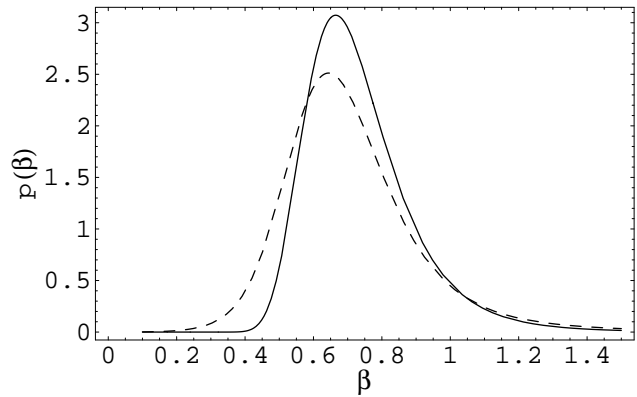}}
\caption{\label{fig:lh_marg} The marginal distribution for $\beta$. 
The result of marginalizing over all possible values of $b$ (from zero
to infinity) is shown as a dashed line. The result of marginalizing
over the values of $b$ in the range $[0.7, 1.1]$ is shown as a solid
line.}
\end{figure}

To do this, one approach is not to use any external information on
bias. In such a case, we marginalize over all possible values of $b$
(from zero to infinity). The resulting distribution for $\beta$ is
shown in Figure~\ref{fig:lh_marg} as a dashed line. The distribution
is positively skewed, with lower limit on $\beta$ being stronger than
the upper one. Specifically, we find that $\beta =
0.64^{+0.24}_{-0.11}$ ($68$\% confidence limits).

A better approach is to use the fact that the square root of the
variance of the PSCz galaxy counts at 8 \hmpc\ is $\s_8^{PSCz} \simeq
0.75$ (Sutherland et al. 1999). Combined with WMAP's estimate of
$\s_8$, this yields for the bias of the PSCz galaxies the value about
$0.9$.\footnote{Our best value of the bias, $0.94$, is, given the
errors, surprisingly close to this estimate.} Therefore, we adopt here
a conservative prior for the bias, namely that it is constrained to
lie in the range $[0.7, 1.1]$. These limits are marked in
Figure~\ref{fig:lh} as dashed vertical lines. We marginalize the
likelihood over the values of $b$ in this range. The resulting
distribution for $\beta$ is shown in Figure~\ref{fig:lh_marg} as a
solid line. This distribution is also skewed and more peaked than the
previous one. We obtain $\beta = 0.66^{+0.21}_{-0.07}$ ($68$\%
confidence limits).

\section{Summary and conclusions}
\label{sec:conc}
We have performed a likelihood analysis of the LG acceleration, paying
particular care to nonlinear effects. We have adopted a widely
accepted assumption that the joint distribution of the LG acceleration
and velocity is Gaussian. Then, two quantities describing nonlinear
effects are relevant. The first one is the coherence function, or the
cross-correlation coefficient of the Fourier components of the gravity
and velocity fields. The second one is the ratio of the power spectrum
of the velocity to the power spectrum of the gravity. Extending our
previous work, we have studied both the coherence function and the
ratio of the power spectra. Using numerical simulations we have
performed fits to the two as functions of the wavevector and
$\s_8$. Then, we have estimated the best values of the parameter
$\beta$ and its errors. We have obtained $\beta =
0.66^{+0.21}_{-0.07}$ at 68\% confidence level.

The analysis of the LG acceleration performed by S92 and S99 was in a
sense more sophisticated than ours. Both teams analyzed a differential
growth of the gravity dipole in subsequent shells around the
LG. Instead, here we used just one measurement of the total
(integrated) gravity within a radius of $150\;\hmpc$. Nevertheless,
the errors on $\beta$ we have obtained are significantly smaller than
those of S92 and S99. In particular, S99 obtained $\beta =
0.70^{+0.35}_{-0.20}$ at $1\s$ confidence level. Comparing these
errors to ours should be done with caution, because S99 considered the
joint likelihood for $\beta$ and the index of the power spectrum,
$\Gamma$.\footnote{The constrains on $\Gamma$ obtained by S99 were
extremely weak, so we decided to fix the value of $\Gamma$ and instead
to study the dependence on $b$.} (To obtain a constraint on $\beta$,
they marginalized the distribution over the values of $\Gamma$ allowed
by the constraints on the Hubble constant.) Still, it is striking that
while our best value of $\beta$ is close to theirs, our errors are
significantly smaller. The reason is our careful modelling of
nonlinear effects. In a previous paper (C02) we showed that the
coherence function used by S99 greatly overestimates actual
decoherence between nonlinear gravity and velocity. Tighter
correlation between the LG gravity and velocity should result in a
smaller random error of $\beta$; in the present work we have shown
this to be indeed the case.

The second factor in the LG acceleration analysis, describing
nonlinear effects, is the ratio of the power spectra of velocity to
gravity. Unlike the coherence function, this quantity has not been
accounted for previously. It affects the value of the correlation
coefficient, hence the random error, to lesser extent than the
CF.\footnote{If we write $\tilde\calR(k) = 1 - \eps(k)$, from
equations~(\ref{eq:err_errors})--(\ref{eq:scaled_err}) it is
straightforward to show that $\err = \err_{|\tilde\calR = 1} +
\calO(\eps^2)$.} However, it does affect the most likely value of
$\beta$, as can be easily noticed from illustrative
equation~(\ref{eq:beta1}) (via $s_\bfv$). Neglecting different
nonlinear growth rates of the gravity and the velocity is equivalent
to setting $\tilde\calR = 1$; we have checked that then the best value
of $\beta$ is $0.62$. Therefore, a small discrepancy between our and
S99's most likely value of $\beta$ is not due to nonlinear effects.

Unlike ours, the analysis of S99 included also other constraints on
the velocity field around the LG: its (small) shear, and the value of
the bulk flow within $30\;\hmpc$. This may have had an effect on the
best value of $\beta$. Moreover, an approach with multiple windows
should allow to further tighten the errors. It is therefore
interesting to repeat the analysis of S99 exactly, but with proper
treatment of nonlinear effects. We plan to do this in the future.

\section*{Acknowledgments}
This research has been supported in part by
the Polish State Committee for Scientific Research grants
No.~2.P03D.014.19 and 2.P03D.017.19. The numerical computations
reported here were performed at the {\em Interdisciplinary Centre for
Mathematical and Computational Modelling}, Pawi\'nskiego 5A,
PL-02-106, Warsaw, Poland.

\end{document}